\documentclass[apj]{emulateapj}
\usepackage{graphicx}

\def\simlt{\mathrel{\hbox{\rlap{\hbox{\lower4pt\hbox{$\sim$}}}\hbox{$<$}}}}
\def\simgt{\mathrel{\hbox{\rlap{\hbox{\lower4pt\hbox{$\sim$}}}\hbox{$>$}}}}
\newcommand{\etal}{et al.}

\def\ro{{\it ROSAT\/}}
\def\chandra{{\it Chandra}}

\def\epks{{\rm 1E~1207.4$-$5209}}

\def\p{\phantom{0}}

\slugcomment{}

\shorttitle{X-ray Observations of Disrupted Recycled Pulsars}
\shortauthors{Gotthelf et al.}

\begin{document}

\title{
X-ray Observations of Disrupted Recycled Pulsars:\\
No Refuge for Orphaned Central Compact Objects
}

\author{E. V. Gotthelf\altaffilmark{1}, J. P. Halpern\altaffilmark{1}, B. Allen\altaffilmark{2,3,4}, B. Knispel\altaffilmark{2,3}}

\altaffiltext{1}{Columbia Astrophysics Laboratory, Columbia University,
550 West 120th Street, New York, NY 10027, USA}
\altaffiltext{2}{Max-Planck-Institut f\"ur Gravitationsphysik, D-30167 Hannover, Germany}
\altaffiltext{3}{Leibniz Universit\"at Hannover, D-30167 Hannover, Germany}
\altaffiltext{4}{Physics Dept., U. of Wisconsin - Milwaukee, Milwaukee WI 53211, USA}

\begin{abstract} 

  We present a \chandra\ X-ray survey of the disrupted recycled
  pulsars (DRPs), isolated radio pulsars with $P>20$~ms and
  $B_s<3\times10^{10}$~G.  These observations were motivated as a
  search for the immediate descendants of the $\approx 10$ central
  compact objects (CCOs) in supernova remnants, three of which have similar
  timing and magnetic properties as the DRPs, but are bright, thermal
  X-ray sources consistent with minimal neutron star cooling curves.
  Since none of the DPRs were detected, there is no evidence that they
  are ``orphaned'' CCOs, neutron stars whose supernova remnants has
  dissipated.  Upper limits on their thermal X-ray luminosities are in
  the range log $L_x[{\rm erg \ s}^{-1}]=31.8-32.8$, which implies cooling
  ages $>10^4-10^5$~yr, roughly 10 times the ages of the $\approx10$
  known CCOs in a similar volume of the Galaxy.  The order of a
  hundred CCO descendants that could be detected by this method are
  thus either intrinsically radio quiet, or occupy a different region
  of ($P,B_s$) parameter space from the DRPs.  This motivates a new
  X-ray search for orphaned CCOs among radio pulsars with larger
  $B$-fields, which could verify the theory that their fields are
  buried by fall-back of supernova ejecta, but quickly regrow to join
  the normal pulsar population.

\end{abstract}


\section {Introduction}
\label{sec:introduction}

The group of about $10$ so-called central compact objects (CCOs) in supernova
remnants (SNRs) are distinguished by their steady surface thermal X-ray
flux, lack of surrounding pulsar wind nebula, and non-detection at any
other wavelength \citep{hal10a}.  Three CCOs are known
pulsars, with periods in the range $0.1-0.4$~s, and spin-down rates that
provide an estimate of their surface dipole magnetic field strength,
which falls in the range $B_s=(3-10)\times10^{10}$~G
\citep{got13}, smaller than that of any other young neutron star (NS).
This weak magnetic field is evidently the physical basis of
the CCO class.

The homogeneous properties of the approximately seven remaining CCOs
that have not yet been seen to pulse suggest that they have similar or
even weaker $B$-fields than the known CCO pulsars, and a more uniform
surface temperature.  That CCOs are found in SNRs (of ages
$300-7000$~yr) in comparable numbers to other classes of NSs implies
that they must represent a significant fraction of NS births, probably
greater than that of magnetars, for example, as only 4--5 Galactic
SNRs are known to host magnetars \citep{hal10b}.

The subsequent evolution of CCOs is a glaring unknown,
their immediate descendants not being evident
in any existing survey.   CCOs should persist as
cooling NSs, detectable in thermal X-rays, for $10^5-10^6$~years
according to NS cooling curves \citep{pag09}.
If some are also radio pulsars,
that phase could last for $\sim10^9-10^{10}$~years.
While there are not yet enough CCOs to know whether they are
intrinsically radio-quiet, it is very unlikely that
the huge expected population of CCO descendants are
{\it all\/} hiding simply due to unfavorable radio beaming.
Therefore, it is difficult to understand why the region
of ($P,\dot P$) space in which CCOs are found, between
the bulk of the ordinary radio pulsars and the recycled
``millisecond''pulsars in binary systems,
is relatively empty.

\begin{figure}[b]
  \includegraphics[width=1.\columnwidth]{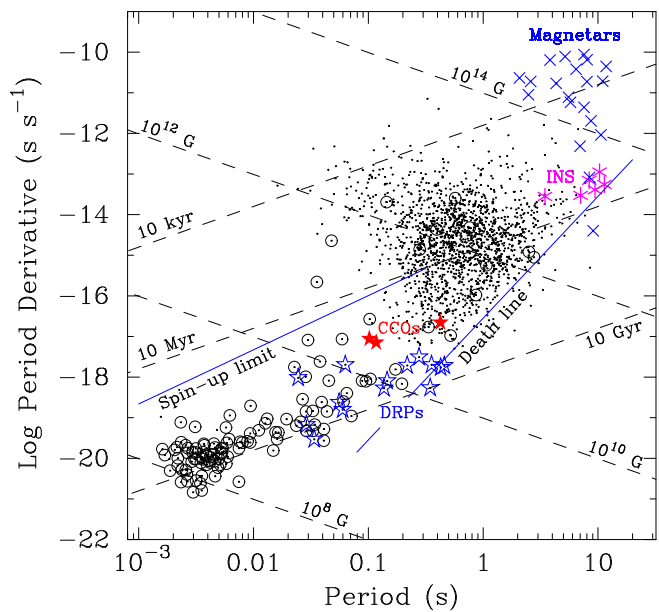}
\caption{Pulsar populations on the $P$-$\dot P$ diagram,
  including magnetars (blue crosses), INSs (magenta asterisks), 
  CCOs (filled red stars) and DRPs (open blue stars).  Black dots are
  isolated pulsars and circled dots are pulsars in binaries.
  (Pulsars in globular clusters are excluded as their period
  derivatives are not entirely intrinsic.)  Dashed lines of
  constant characteristic age and magnetic field are indicated.
  }
\label{fig:ppdot}
\end{figure}

Most of the pulsars in this sparse region (see Figure~\ref{fig:ppdot})
are thought to be ``mildly recycled,'' having been spun up by accretion
from a high-mass companion for a relatively short time
before a second SN occurred.
Defined as having $P>20$~ms and $B_s<3\times10^{10}$~G,
mildly recycled pulsars include double NS systems,
and single ones thought to be the disrupted
recycled pulsars (DRPs, \citealt{lor04}) ejected when
the binary is unbound after the second SN.
(These are in contrast to the millisecond
pulsars, which have low-mass companions.)

The DRPs have characteristic ages $\tau_c\equiv P/2\dot P$
of $10^9-10^{10}$~yr.
Historically, it was thought that hardly any pulsars are born
with $B_s<10^{11}$~G, so that all such pulsars must be recycled.
But the discovery of young CCOs in this region of parameter space
invalidates that assumption.
Just as the $\sim10^8$~yr characteristic age of a CCO is meaningless,
the possibility that {\it any} low $B$-field radio pulsar
is much younger than its characteristic age may now be considered.

\begin{deluxetable*}{lllrrclrccc}
\tabletypesize{\scriptsize}
\tablewidth{0pt}
\tablecaption{Properties of Disrupted Recycled Pulsars}
\tablehead{
\colhead{PSR} & \colhead{R.A. (J2000)}  &\colhead{Decl. (J2000)} & \colhead{$P$}  &\colhead{$\dot P$}    &\colhead{DM}   &\colhead{ $d_{\rm DM}$\tablenotemark{a}}   & \colhead{$|z|$} &\colhead{$B_s$} &\colhead{$\dot E$} & Reference  \\
\colhead{Name}& \colhead{(h\ \ m\ \ s)} &\colhead{($^{\circ}\ \ ^{\prime}\ \ ^{\prime\prime}$)} & \colhead{(ms)} &\colhead{ } &\colhead{(cm$^{-3}$ pc)}& \colhead{(kpc)}  &\colhead{(pc)}  &\colhead{(G)} &\colhead{(erg s$^{-1}$)} 
}
\startdata
J0609$+$2130 & 06 09 58.89 & $+$21 30 02.8 &   56 & $2.35\times10^{-19}$ &  38.73 & 1.2  &   22 & $3.66\times10^{9}$\phantom{0} & $5.4\times10^{31}$ & \p1 \\
J1038$+$0032 & 10 38 26.93 & $+$00 32 43.6 &   29 & $6.70\times10^{-20}$ &  26.59 & 1.2  &   880  & $1.41\times10^{9}$\phantom{0} & $1.1\times10^{32}$ & \p2 \\
J1320$-$3512 & 13 20 12.68 & $-$35 12 26.0 &  458 & $1.9\times10^{-18}$ &  16.42 & 0.68  &   310  & $2.99\times10^{10}$ & $7.8\times10^{29}$ & \p3 \\
J1333$-$4449 & 13 33 44.83 & $-$44 49 26.2 &  346 & $5.4\times10^{-19}$ &  44.3 & 1.4  &   410  & $1.38\times10^{10}$ & $5.2\times10^{29}$ & \p4 \\
J1339$-$4712 & 13 39 56.59 & $-$47 12 05.5 &  137 & $5.3\times10^{-19}$ &  39.9 & 1.2  &   310  & $8.62\times10^{9}$\phantom{0} & $8.1\times10^{30}$ & \p4 \\
J1355$-$6206 & 13 55 21.34 & $-$62 06 20.1 &  277 & $3.1\times10^{-18}$ & 547 & 8.3  &   22 & $2.96\times10^{10}$ & $5.8\times10^{30}$ & \p5\\
J1548$-$4821 & 15 48 23.26 & $-$48 21 49.7 &  146 & $8\times10^{-19}$ & 126.0 & 4.4  &   360  & $1.09\times10^{10}$ & $1.0\times10^{31}$ & \p5 \\
J1611$-$5847 & 16 11 51.31 & $-$58 47 42.3 &  355 & $2.0\times10^{-18}$ &  79.9 & 1.7  &   160  & $2.70\times10^{10}$ & $1.8\times10^{30}$ & \p6 \\
J1753$-$1914 & 17 53 35.17 & $-$19 14 58 &     63 & $2.02\times10^{-18}$ & 105.3 & 2.2  &   130  & $1.14\times10^{10}$ & $3.2\times10^{32}$ & \p6 \\
J1816$-$5643 & 18 16 36.46 & $-$56 43 42.1 &  218 & $1.93\times10^{-18}$ &  52.4 & 1.6  &   470  & $2.08\times10^{10}$ & $7.4\times10^{30}$ & \p4 \\
J1821$+$0155\tablenotemark{b} & 18 21 38.88 & $+$01 55 22.0 &   34 & $2.94\times10^{-20}$ &  51.75 & 1.8 &  24  & $1.01\times10^{9}$\phantom{0} & $3.0\times10^{31}$ & \p7 \\
B1952$+$29\p & 19 54 22.55 & $+$29 23 17.3 &  427 & $1.71\times10^{-18}$ &   7.932 & 0.70  &   9 & $2.73\times10^{10}$ & $8.7\times10^{29}$ & \p8 \\
J2007$+$2722 & 20 07 15.83 & $+$27 22 47.91 &   24 & $9.61\times10^{-19}$ & 127.0 & 5.4  &   250  & $4.91\times10^{9}$\phantom{0} & $2.6\times10^{33}$ & \p9 \\
J2235$+$1506 & 22 35 43.70 & $+$15 06 49.1 &   60 & $1.58\times10^{-19}$ &  18.09 & 1.1  &   630  & $3.11\times10^{9}$\phantom{0} & $2.9\times10^{31}$ & \hfill 10\p
\enddata
\tablerefs{
(1) \citealt{lor05};
(2) \citealt{bur06};
(3) \citealt{dam98};
(4) \citealt{jac09};
(5) \citealt{kra03};
(6) \citealt{lor06};
(7) \citealt{ros12};
(8) \citealt{hob04};
(9) \citealt{all13};
(10) \citealt{cam96}.
}
\tablenotetext{a}{DM distance derived using the NE2001 Galactic free electron density model of \citet{cor02}.}
\tablenotetext{b}{PSR J1821+0155 was discovered too recently to be included in this X-ray study.}
\label{tab:drps}
\end{deluxetable*}

The majority of CCOs may have magnetic fields even weaker than those
of the known CCO pulsars, and may fall among the DRPs in ($P,\dot P$)
space.  Once the SNR associated with a CCO has dissipated, it would be
difficult to distinguish an ``orphaned CCO'' from a DRP by timing
alone if some CCOs are radio pulsars.  Thermal X-ray emission,
however, would allow a recently orphaned CCO to be recognized as such
up to $\sim10^6$~yr.  Thermal emission from the cooling NS is the
diagnostic that would distinguish an evolving CCO from an old DRP,
whose negligible rotation-powered X-ray emission, thermal or
non-thermal, would be orders of magnitude weaker.

In this paper, we report an X-ray search for orphaned CCOs from among
the population DRPs, whose timing parameters are expected to be
comparable.  In Section~\ref{sec:observations} we describe the new and
archival \chandra\ observations of the DRPs. Section~\ref{sec:results}
gives the resulting upper limits on their temperatures and
luminosities.  In Section~\ref{sec:discussion} we discuss the
implication of these results for the possible evolutionary tracks of
CCOs.

\section{Observations}
\label{sec:observations}

Our targets selected for X-ray observations are the 12 radio pulsars
classified as DRPs by \cite{bel10}, plus the recently discovered
PSR~J2007+2722 \citep{kni10}. These comprise all but one of the
isolated pulsars in the Galactic disk with magnetic field strength
$B_s< 3\times 10^{10}$~G and spin period $P>20$~ms listed in the ATNF
catalog\footnote{http://www.atnf.csiro.au/research/pulsar/psrcat/expert.html}
\citep[v1.46]{man05}.  Their properties are listed in
Table~\ref{tab:drps}.  The latest DRP, PSR~J1821+0155 \citep{ros12},
the $14^{\rm th}$ member of the class,  was discovered too recently to
be included in our X-ray sample.

For ten of these objects not already observed in X-rays, we
obtained 3.5~ks \chandra\ observations to search for point-like
emission at their known (subarcsecond) radio locations.
We justified this short observing time based on its ability to
detect thermal emission from a cooling NS younger
than $\sim10^5$~yr, while thermal or nonthermal emission
from a $\sim10^9$~yr old DRP would be many orders of magnitude less.
Detailed calculations of the detection limits on temperature
and luminosity from these observations are presented below.
We also analyzed 5~ks archival exposures on PSR~J0609$+$2130
and PSR~B1952$+$29, and tabulate our prior results for
PSR~J2007$+$2722 \citep{all13}.

All observations were taken with the Advanced CCD Imaging Spectrometer
\citep[ACIS,][]{gar03}, operating in timed/faint exposure mode,
with the targets placed on the back-illuminated \hbox{ACIS-S3} CCD.
ACIS has $0.\!^{\prime\prime}5$ pixels, comparable to the on-axis
point-spread function. The nominal ACIS  pointing uncertainty is
a radius of $0.\!^{\prime\prime}6$. All data reduction and analysis was performed
with the \chandra\ Interactive Analysis of Observation software
\citep[CIAO,][]{fru06}
version 4.5, using the calibration database (CALDB) v4.1.3. The
background rates for these observations showed no evidence of flaring
behavior, and the full exposure time was retained for analysis in each case.

\begin{deluxetable*}{cccccl}
\tabletypesize{\scriptsize}
\tablewidth{0pt}
\tablecaption{Upper Limits on X-ray Emission from DRPs}
\tablehead{
\colhead{PSR Name} & \colhead{\chandra} &\colhead{Livetime} &\colhead{$N_{\rm H}^{\rm DM}$~\tablenotemark{a}}&\colhead{$kT_{\rm max}$\tablenotemark{b}} &\colhead{$L^{\rm bol}_{\rm max}$\tablenotemark{b}} \\
\colhead{}& \colhead{ObsID}      &\colhead{(ks)}       &\colhead{(cm$^{-2}$)}  &\colhead{(eV)}    &\colhead{(erg s$^{-1}$)} 
}
\startdata
J0609$+$2130 & 12687 &   4.99 &     $1.2\times10^{21}$   & $49$  & $1.6\times 10^{32}$ \\
J1038$+$0032 & 13801 &   3.50 &     $8.2\times10^{20}$   & $50$  & $1.6\times 10^{32}$ \\
J1320$-$3512 & 13797 &   3.42 &     $5.1\times10^{20}$   & $43$  & $9.4\times 10^{31}$ \\
J1333$-$4449 & 13800 &   3.42 &     $1.4\times10^{21}$   & $54$  & $2.4\times 10^{32}$ \\
J1339$-$4712 & 13799 &   3.42 &     $1.2\times10^{21}$   & $52$  & $2.0\times 10^{32}$ \\
J1355$-$6206 & 13806 &   3.42 &     $1.7\times10^{22}$   & $135$ & $9.0\times 10^{33}$ \\
J1548$-$4821 & 13805 &   3.42 &     $3.9\times10^{21}$   & $84$  & $1.3\times 10^{33}$ \\
J1611$-$5847 & 13802 &   3.41 &     $2.5\times10^{21}$   & $62$  & $4.1\times 10^{32}$ \\
J1753$-$1914 & 13803 &   3.42 &     $3.3\times10^{21}$   & $69$  & $6.2\times 10^{32}$ \\
J1816$-$5643 & 13804 &   3.42 &     $1.6\times10^{21}$   & $57$  & $2.9\times 10^{32}$ \\
B1952$+$29\p & 12684 &   4.99 &     $2.5\times10^{20}$   & $40$  & $6.9\times 10^{31}$ \\
J2007$+$2722&6438,7254,8492&94.04&  $3.9\times10^{21}$   & $68$  & $5.8\times 10^{32}$ \\
J2235$+$1506 & 13798 &   3.42 &     $5.6\times10^{20}$   & $47$  & $1.4\times 10^{32}$
\enddata
\tablenotetext{a}{$N_{\rm H}^{\rm DM}$ approximated as $10\times$DM.
}  
\tablenotetext{b}{Upper limit on blackbody temperature and bolometric luminosity
  at 99.73\% confidence for a cooling NS of radius $R_{\infty} = 14.5$~km
  at the DM distance given in Table~\ref{tab:drps}.}  
\label{tab:limits}
\end{deluxetable*}

\section{Results}
\label{sec:results}

\begin{figure*}
\centerline{
\hfill
 \includegraphics[height=1.2\linewidth]{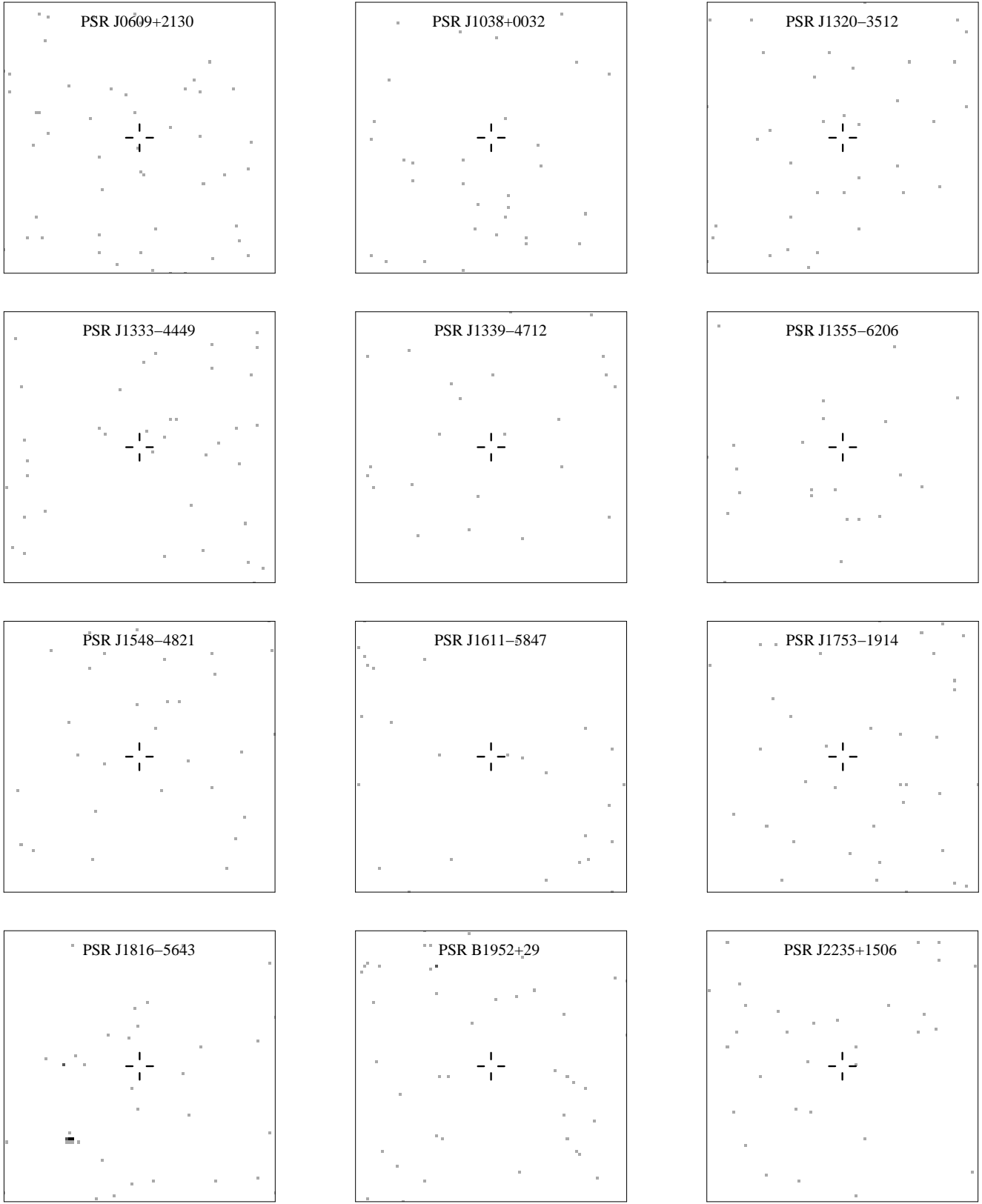}
\hfill
}
\vspace{0.2in}
\caption{\chandra\ ACIS-S $0.3-10$~keV X-ray images of targeted radio
  pulsars listed in Table~\ref{tab:drps}. Each grey square is a
  detected count and black squares contain two counts. A nearby source to
  PSR~J1816-5643 is evident by a cluster of counts.  No X-ray sources
  are found within the adopted $1.\!^{\prime\prime}2$ radius \chandra\
  error circle centered on the radio coordinates (crosses). The plots
  are $45^{\prime\prime}$ on a side; the inner dimensions of the crosses
  are $2^{\prime\prime}$.}
\label{fig:images}
\end{figure*}

Figure \ref{fig:images} presents \chandra\ thumbnail images in
the 0.3--10~keV band centered around the radio coordinates of
each DRP, excluding PSR~J2007$+$2722 reported elsewhere \citep{all13}.
In these $45^{\prime\prime}\times45^{\prime\prime}$ sub-images, no pixel that
is not definitely associated with a significant source
contains more than two counts in the $0.3-10$~keV energy band. 
Examination of each image shows no evidence for a source at the radio
location within twice the nominal $r=0.\!^{\prime\prime}6$ pointing uncertainty.
In fact, no counts are detected in an adopted aperture of
radius $1.\!^{\prime\prime}2$ at the position of any target.
This is not unexpected given the mean
background rate of $\approx1.1\times10^{-6}$ counts~s$^{-1}$~pixel$^{-1}$,
uniform across the 12 observations.  For this rate,
the mean number of counts in a 3.5~ks observation
is 0.073 in a $1.\!^{\prime\prime}2$ radius circle.
There is a $93\%$ probability of detecting no counts in
that aperture for a single observation,
and only a $58\%$ chance of getting one or more
counts in any of 12 observations. In no case are the coordinates
of the nearest detected X-ray source consistent with the radio location,
the closest being $\approx16^{\prime\prime}$ from PSR~J1816$-$5643.

With no evidence of any photon at the location of each DRP, we
calculate an upper limit on the thermal flux from an assumed cooling
NS of radius $R_{\infty}=14.5$~km, to match the radius used to derive
the theoretical cooling curves discussed in Section
\ref{sec:discussion}. As photon counts follow the Poisson
distribution, the probability of having gotten zero photons is 0.0023
when the expected number of photons from a source is six.  Therefore
the $99.77\%$ confidence ($3\sigma$) upper limit on the source flux is
that which would predict six counts.  We determine the blackbody
temperature required for a fiducial source to produce six counts plus
background in the detector by convolving an absorbed blackbody
spectrum through the ACIS spectral response and computing the total
counts in the $0.3-10$~keV bandpass generated for each observation. The
blackbody flux normalization is fixed by the ratio $(R_{\infty}/d_{\rm
  DM})^2$ for each target, where $d_{\rm DM}$ is the distance derived
using the NE2001 Galactic free electron density model of
\citet{cor02}.  An absorbing column density $N_{\rm H}^{\rm DM}$ is
estimated from the dispersion measure (DM) assuming a rule-of-thumb
$N_{\rm H}^{\rm DM}/N_{\rm e} \approx 10$, i.e., $N_{\rm H}^{\rm
  DM}=10\times$DM \citep{he13}.  Table~\ref{tab:limits} presents the
upper limits computed in this way on the blackbody temperature and
bolometric luminosity of each pulsar, quantities measured at infinity.
These generally correspond to $T_{\rm max}$ in the range
$(5-8)\times10^5$~K and log $L_{\rm max}^{\rm bol}[{\rm erg
    \ s}^{-1}]=31.8-32.8$.  The two outliers are PSR~J1355$-$6206 and
PSR~J1548$-$4821, which are less constrained because of their large
distances and $N_{\rm H}^{\rm DM}$.

The uncertainties on these upper limits are dominated by systematic
errors involving the DM derived distances and column densities.
DM distance can have fractional
uncertainty of 25\% or larger \citep[e.g., ][]{cam09}.
The neutral column density estimated using a typical
ionized fraction involves another uncertain assumption.
Furthermore, an error on $N_{\rm H}$ amplifies the error on the
temperature measurement, which comes from the low-energy end of
the ACIS-S instrument response, around $0.3$~keV, where the
detector sensitivity falls off rapidly and is poorly calibrated.
Unfortunately, these effects are difficult to quantify.

We repeat, for completeness, that we would not expect to detect any
of the DRPs if they are old, rotation-powered NSs with spin-down power
$\dot E$.  For comparison, we can use the dozen old pulsars whose
X-ray detections were compiled by \citet{pos12a}.  These typically have
$L_x(1-10\,{\rm keV})\sim10^{-3}\,\dot E$ with a scatter of a factor
of 10.  The same X-ray efficiency for the DRPs would produce
$L_x\sim10^{27}-10^{30}$ erg~s$^{-1}$, which is orders of magnitude
below our upper limits.

\section{Discussion}
\label{sec:discussion}

The upper limits on temperature and luminosity of each
DRP can be compared with standard (minimal) NS cooling curves,
\citep[e.g.,][]{pag09} to place a lower limit on its age.
These limits depend strongly on uncertain variables
such as the critical temperature
for superfluid neutron pairing,
and the composition of the NS envelope, which is
why there cannot be a unique age limit for each entry in
Table~\ref{tab:limits}.  Roughly speaking, a luminosity
limit of log $L_{\rm max}[{\rm erg \ s}^{-1}]=32.8$ requires an
age $\tau>10^4$~yr for heavy element envelopes and
$\tau>3\times 10^4$~yr for light elements, 
while log $L_{\rm max}[{\rm erg \ s}^{-1}]=31.8$
implies that $\tau>5\times10^4$~yr (light) or
$\tau>2\times10^5$~yr (heavy).  The cooling curves for
light and heavy element envelopes cross over in this range
of luminosities.
The upper limits on temperatures and luminosities for the DRPs
(with the possible exception of PSR J1355$-$6206)
are smaller than those of all CCOs but one.
In no case does a DRP overlap in possible age with the SNR
ages of the known CCOs, which are $300-7000$~yr.
The dozen DRPs fail to qualify as evolved CCOs in the age range
that is, roughly speaking, 10 times the ages of the
known CCOs, where we expect their descendants to be 10
times as numerous.

The meaning of these X-ray non-detections of DRPs
for the evolution of CCOs depends on the volume
sampled by the surveys that discovered both populations,
and their relative completeness.  Both are
difficult to evaluate; however, the volumes appear to be
at least comparable.  The $\approx 10$ CCOs are found in SNRs
up to a maximum distance of $\simlt 8$~kpc, and the DRPs
appear to have a similar distribution of distance and
Galactic coordinates.  Therefore, the absence of radio
pulsar counterparts of orphaned CCOs appears to be real,
at least in the range of magnetic field strengths which
define the DRPs.  \citet{bel10} noted that roughly four
of the DRPs so defined could actually be interlopers
from the population of normal pulsars, as extrapolated
from the statistics of studies such as \citet{fau06}.
However, as we argued previously, it may not be
possible to make such a distinction.  In any case,
it would not change our conclusions regarding
the fate of CCOs, that there are no known radio pulsars
with $B_s<3\times10^{10}$~G that are their immediate,
$\tau<10^5$~yr old, descendants, where we would expect
to find $\sim10^2$ orphans.

Another clue to the age of DRPs should be their distribution
of heights $z$ above the Galactic plane as listed in
Table~\ref{tab:drps}.  However, as discussed by \citet{bel10},
these heights are smaller than one would expect
for the average NS kick velocity of $265$~km~s$^{-1}$ \citep{hob05},
which makes it difficult to use $z$ as an indicator of age
for DRPs.  At this velocity a NS would travel only 270~pc
in $10^6$~yr, implying that X-ray detected orphaned CCOs
could have a similar $z$ height as the DRPs, which are thought
to be much older.  Since they are old, the small scale
height of the DRPs still requires an explanation.
\citet{bel10} propose that the first SN in the parent binary
was of a different type that would give little or
no kick to the system, perhaps an electron-capture SN.

A priori, one might not have expected DPRs to be orphaned
CCOs.  As it is, there are not enough DRPs compared to
double NS systems according to standard evolutionary models that
link them \citep{bel10}.
Any DRP that is reassigned to a different population would
only exacerbate this shortage.  Still, the evolutionary
fate of CCOs remains unknown after this survey.

One possible solution is that radio luminosity is a declining function
of spin-down power.  If so, radio surveys could be grossly
incomplete in detecting such low $\dot E$ pulsars even though they are
on the active side of the radio pulsar death line.  There is good
evidence that ordinary radio pulsars behave this way, with $
L_r \propto \dot E^{1/2}$ \citep{fau06}, because there is no pileup in the
number of pulsars near the death line.  However, it is not clear that
this effect alone could explain the absence of orphaned CCOs, because
there are in fact many radio pulsars with lower $\dot E$ than the CCO
pulsars. Such an effect may also apply to the seven \ro\ discovered,
radio-quiet isolated neutron stars \citep[INSs,][]{hab07} which,
however, have strong magnetic fields \citep{kap09}, but are close to
the radio pulsar death line.  The INSs (Fig.~\ref{fig:ppdot}) are a
good analogy to our problem in that they {\it are} plausibly the
descendants of the magnetars, following a fast epoch of magnetic field
decay around $\sim10^4$~yr \citep{col00}.  It is likely that the INSs
are kept hot for longer than CCOs by their continuing magnetic field
decay for up to $\sim10^6$~yr \citep{pon07}, which could account for
their abundance relative to the elusive orphaned CCOs.

It may be difficult to detect and/or
recognize orphaned CCOs if they cool faster than ordinary
NSs.  One effect that can accelerate cooling
is an accreted light-element envelope, which has higher
heat conductivity than an iron surface \citep{kam06}.
However, this effect actually makes CCOs hotter than bare NSs
for their first $10^5$~yr, after which their temperatures
plummet.  Therefore, the prediction that CCO descendants
should be detectable in soft X-rays remains robust.

Another plausible home for orphaned CCOs would be among
the radio pulsars with magnetic fields comparable to or
higher than those of the CCO pulsars.  
One theory for CCOs postulates
that they are born with a canonical NS magnetic field of
$\sim10^{12}$~G that was largely buried by fall-back of a small
amount of supernova ejecta, $\sim10^{-5}-10^{-4}\,M_{\odot}$, during
the hours and days after the explosion. The buried
field will diffuse back to the surface on a time scale that is
highly dependent on the amount of mass accreted
\citep{mus95,ho11,vig12,ber12},
after which the CCOs will join the bulk of the population of ordinary
pulsars.  For accretion of $\sim10^{-5}\,M_{\odot}$,
the regrowth of the surface field is largely complete after $\sim10^3$~yr,
but if $>0.01\,M_{\odot}$ is accreted, then the
diffusion time could be millions of years.

Such a scenario addresses the absence of CCOs descendants; they
turn into ordinary pulsars.  It also has the advantage of not requiring
yet another class of NS to exist that would only exacerbate the
apparent excess of pulsars with respect to the Galactic core-collapse
supernova rate, a problem emphasized by \citet{kea08}.
Furthermore, magnetic field growth has long been considered a reason
why measured pulsar braking indices are all less than the dipole value
of 3.  In this picture, CCOs represent one extreme in the evolution of
surface magnetic field, and almost {any\/} radio pulsar {\it might\/}
be a former CCO.  Finally, an intrinsically strong crustal magnetic
field appears to be necessary to explain the existence of the thermal
hot spots that enable us to detect pulsations from CCOs in the first
place (see discussion in \citealt{got13}).

For the first $\sim10^5$~yr, rapid field growth can only move a CCO
vertically upward in the $P-\dot P$ diagram.  Such movement is
difficult to detect directly using CCOs, because it would require
measuring the braking index or observing the change of the dipole
magnetic field spectroscopically, neither of which is likely to be
possible if the relevant time scale is $\ge10^3$~yr.  However, during
their first $10^5$~yr, orphaned CCOs in this scenario should still
have periods of $\sim0.1-0.5$~s and could have magnetic fields in the
range $3\times10^{10}-3\times10^{11}$~G.  A search of all 159 isolated
radio pulsars in this range for thermal X-ray emission from such
``old'' pulsars would provide a promising avenue for finding orphaned
CCOs. Of these, two are known X-ray sources, the faint ($\sim
10^{29}$~erg~s$^{-1}$) nearby radio pulsars PSR B1451$-$68 and PSR
B0950+08. X-rays from these sources are attributed to a combination of
heated polar caps and non-thermal (magnetospheric) emission
\citep{pos12b,zav04}. If further X-ray surveys of radio pulsars fail to
find any orphaned CCOs, then it will be difficult to escape the
conclusion that they are intrinsically radio quiet.

\section{Conclusions}

Following the discovery that CCOs have weaker magnetic fields
than any other young pulsar, it became apparent that their
descendants were not obviously present in radio or X-ray surveys.
If their magnetic fields at birth are intrinsic, and do not
change with time, then the region around the CCOs in the
$(P,\dot P )$ diagram of radio pulsars should be densely populated
with all of their descendants, unless they are radio quiet.
The fact that this area is quite sparsely populated led us to survey
a large fraction of the available radio candidates in X-rays,
those which were previously understood to be mildly recycled pulsars.
The ``smoking pulsar'' evidence of an orphaned CCO should
 be an X-ray hot NS that could be detected, in a short observation,
at an age up to $10^5$~yr, which is much younger than the characteristic
ages of the targeted DRPs but much older than the known CCOs.
Only upper limits on their thermal X-ray luminosities were found,
in the range log $L_x[{\rm erg \ s}^{-1}]=31.8-32.8$, which implies cooling
ages $>10^4-10^5$~yr.

Up to the age limits implied by the X-ray
non-detections, there should be $\sim100$ CCO descendants in
the volume sampled by radio pulsar surveys.  Since none have
been found among radio pulsars with $B_s<3\times10^{10}$~G,
the next step should be to search for young, cooling NSs among
the radio pulsars with larger $B$-fields, comparable to
or even larger than that of the CCO \epks, with $B_s=1\times10^{11}$~G.
An especially interesting possibility is that CCOs have intrinsically
strong $B$-fields that were promptly buried by a small amount of supernova
debris, but will grow back to ``normal'' strength in $\sim10^4$~yr.
If such descendants of CCOs are found in thermal X-rays among
the ordinary radio pulsar population, it would help solve
problems about their surface thermal patterns in addition to their evolution.
Otherwise, if the orphaned CCOs are truly radio silent for some
unknown reason, they could still be found in more sensitive
all-sky surveys in soft X-rays, by analog with the (evidently more
luminous) INSs that were discovered this way.

\acknowledgments

We thank Dr. Reinhard Prix for discussion during the proposal phase.
This investigation is based on observations obtained
with the \chandra\ Observatory.  Financial support was provided by
award GO2-13097X issued by the \chandra\ X-ray Observatory Center,
which is operated by the Smithsonian Astrophysical Observatory for and
on behalf of NASA under contract NAS8-03060.

\end{document}